%% file: main.tex
\documentclass[10pt,twocolumn,letterpaper]{article}

\usepackage[pagenumbers]{iccv} 


\definecolor{iccvblue}{rgb}{0.21,0.49,0.74}
\usepackage[pagebackref,breaklinks,colorlinks,allcolors=iccvblue]{hyperref}
\usepackage{bm}
\usepackage{multirow}
\usepackage[dvipsnames]{xcolor}
\usepackage{color, colortbl}
\definecolor{Gray}{gray}{0.93}
\newcommand{\beginsupplement}{%
        \setcounter{table}{0}
        \renewcommand{\thetable}{S\arabic{table}}%
        \setcounter{figure}{0}
        \renewcommand{\thefigure}{S\arabic{figure}}%
    }
\def\paperID{9878} 
\def\confName{ICCV}
\def\confYear{2025}

\title{Text2CT: Towards 3D CT Volume Generation from Free-text Descriptions Using Diffusion Model}

\author{
{Pengfei Guo$^{1}$ \qquad  Can Zhao$^{1}$  \qquad Dong Yang$^{1}$ \qquad  Yufan He$^{1}$} \\
{Vishwesh Nath$^{1}$ \qquad  Ziyue Xu$^{1}$ \qquad  Pedro R. A. S. Bassi$^{2}$ \qquad  Zongwei Zhou$^{2}$} \\ {\qquad Benjamin Simon$^{3}$ \qquad Stephanie Harmon$^{3}$ \qquad Baris Turkbey$^{3}$ \qquad Daguang Xu$^{1}$}\\
\vspace{0.4em} 
{$^{1}$NVIDIA \quad $^{2}$Johns Hopkins University \quad $^{3}$National Institutes of Health}\\
\vspace{0.4em} 
}

\begin{document}

\maketitle
\input{sec/0_abstract}    
\input{sec/1_intro}
\input{sec/2_related}
\input{sec/3_method}
\input{sec/4_experiments}
\input{sec/5_discussion}
\input{sec/6_conclusion}

{
    \small
    \bibliographystyle{ieeenat_fullname}
    \bibliography{main}
}

\input{sec/7_suppl}

\end{document}

%% file: sec/0_abstract.tex
\begin{abstract}
Generating 3D CT volumes from descriptive free-text inputs presents a transformative opportunity in diagnostics and research. In this paper, we introduce \textbf{Text2CT}, a novel approach for synthesizing 3D CT volumes from textual descriptions using the diffusion model. Unlike previous methods that rely on fixed-format text input, \textbf{Text2CT} employs a novel prompt formulation that enables generation from diverse, free-text descriptions. The proposed framework encodes medical text into latent representations and decodes them into high-resolution 3D CT scans, effectively bridging the gap between semantic text inputs and detailed volumetric representations in a unified 3D framework. Our method demonstrates superior performance in preserving anatomical fidelity and capturing intricate structures as described in the input text. Extensive evaluations show that our approach achieves state-of-the-art results, offering promising potential applications in diagnostics, and data augmentation.  
\end{abstract}

%% file: sec/1_intro.tex
\section{Introduction}\label{sec:intro}

\begin{figure}[t] 
\centering
\includegraphics[width=\columnwidth]{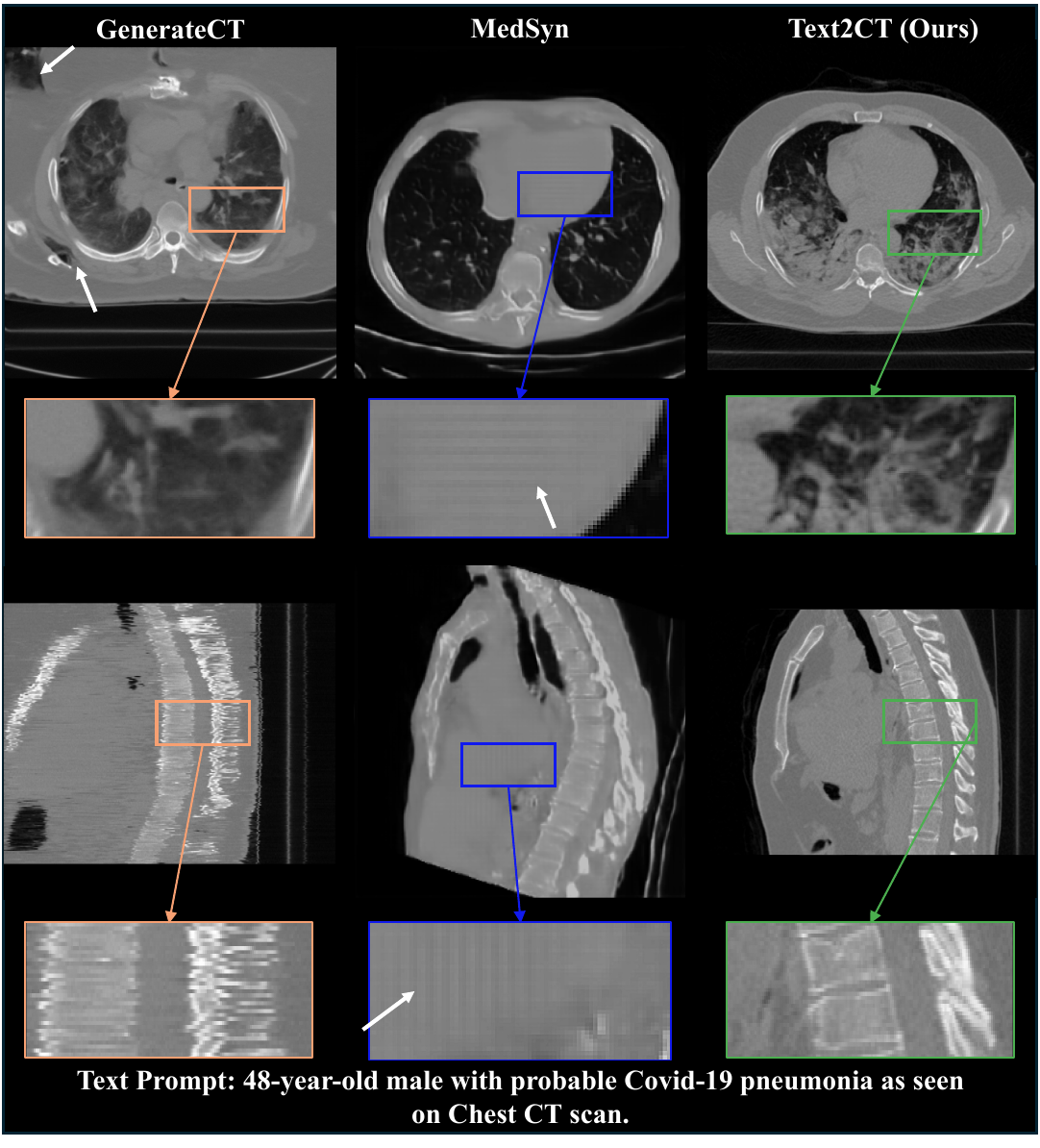}
\caption{Generated 3D CT volume by the proposed Text2CT (left) and GenerateCT~\cite{hamamci2023generatect} (right). We show the axial and sagittal views from top to bottom, respectively. 3$\times$ zoomed-in images are presented to emphasize the differences, highlighting the 3D continuity issue of using a 2D super-resolution network in GenerateCT~\cite{hamamci2023generatect} and the grid-like artifacts from 3D super-resolution network in MedSyn~\cite{xu2024medsyn}. White arrows point to areas of abnormal human anatomy and artifacts from super-resolution in ~\cite{hamamci2023generatect,xu2024medsyn}, revealing potential inaccuracies in the generated output.}
\label{fig:compare_gct}
\end{figure}

Text-to-image generation~\cite{qiao2019mirrorgan, ramesh2021zero, li2019controllable} has become a transformative tool in natural image processing, enabling the synthesis of detailed images from descriptive language. This technology has unlocked new possibilities in areas such as creative media~\cite{oppenlaender2022creativity}, virtual environments~\cite{gafni2022make}, and accessibility tools~\cite{zhang2023text} by allowing users to visualize concepts directly from text inputs. The ability to generate accurate and contextually rich images from text has not only enhanced user interaction but also pushed the boundaries of artificial intelligence in understanding and replicating complex visual scenes. As the technology progresses, it faces the formidable challenge of extending beyond the generation of two-dimensional natural scenes to the more complex domain of three-dimensional medical imaging. The synthesis of 3D medical images from textual descriptions presents a unique set of challenges, including higher demands for accuracy, anatomical precision, and the need for medically relevant imagery that adheres to clinical standards~\cite{shin2018medical,guo2024maisi,frangi2018simulation,yang2023image}. Addressing these challenges not only expands the scope of text-to-image technology but also holds the potential to revolutionize medical diagnostics and personalized treatment planning through tailored, descriptive visualizations of patient-specific conditions.

Medical imaging analysis plays a key role in modern healthcare, with deep learning (DL) models enhancing diagnostic and treatment workflows~\cite{litjens2016deep,alyasseri2022review,mansouri2021deep}. However, critical challenges remain. \textbf{First}, data scarcity limits model generalizability, particularly for rare conditions~\cite{bansal2022systematic,Guo_2021_CVPR}. \textbf{Second}, privacy concerns arise from handling sensitive patient data under strict ethical and regulatory requirements~\cite{kaissis2021end,razzak2018deep,paul2023digitization}. \textbf{Third}, the high cost of expert annotations—required to identify subtle diagnostic features in medical images—presents a persistent barrier~\cite{lutnick2019integrated,tajbakhsh2021guest}. These challenges collectively form significant barriers to developing DL models in the medical field, necessitating innovative solutions to overcome them and fully harness the capabilities of AI-driven medical imaging analysis. To address these challenges,
the generation of synthetic medical data has emerged as a promising solution~\cite{singh2021medical, kazerouni2023diffusion}. Synthetic data offers a way to augment limited datasets, reduce reliance on real patient data, and mitigate the high costs associated with manual annotation~\cite{chlap2021review,garcea2023data}. By creating artificial yet realistic medical images, synthetic data not only enriches existing datasets but also provides a scalable and ethical alternative for training deep learning models.

\begin{figure}[t] 
\centering
\includegraphics[width=\columnwidth]{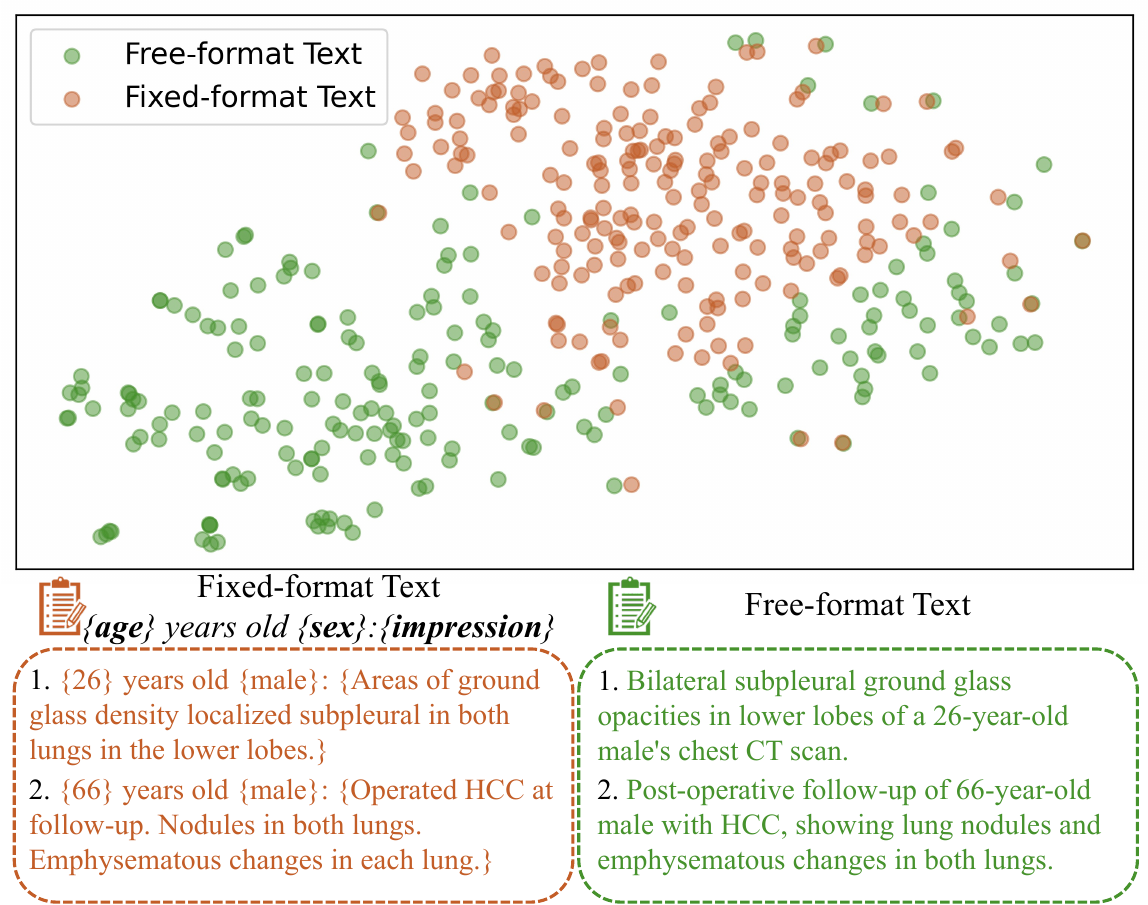}
\caption{\textit{\textbf{Top row}}: t-SNE plot comparing latent feature distributions from free-format and fixed-format texts. The broader spread of free-format features highlights greater diversity and variability, contrasting with the more constrained distribution of fixed-format text. \textit{\textbf{Bottom row}}: Two examples describing the same content using free-format and fixed-format text prompts.
}
\label{fig:tsne_text}
\end{figure}

Despite notable advancements in text-conditional medical image generation~\cite{hamamci2023generatect,xu2024medsyn, bluethgen2024vision}, several critical challenges remain insufficiently addressed in existing research. \textbf{First}, the generation of realistic, high-resolution 3D volumes. This task is particularly challenging due to the immense memory demands of existing 3D frameworks, which must handle the substantial data required for high-dimensional representations~\cite{singh20203d}. To mitigate these memory constraints, some methods~\cite{hamamci2023generatect,zhang2022bridging,chen20242} adopt hybrid architectures that combine 2D and 3D components. However, similar to issues encountered in long video generation, relying on 2D networks~\cite{hamamci2023generatect} can cause continuity problems in generated 3D medical volumes, while 3D two-stage super-resolution~\cite{xu2024medsyn} tends to introduce grid-like artifacts, as illustrated in Fig.~\ref{fig:compare_gct}. Addressing these memory limitations of the unified 3D network is crucial to enhance the realism and clinical applicability of 3D volume generation, where fidelity, precision, and consistency are essential. 
\textbf{Second}, the existing text-conditional medical image generation models impose rigid constraints on the format of input textual descriptions. For instance, previous approaches~\cite{xu2024medsyn, hamamci2023generatect}  limit inputs to predefined structures, such as combining demographic information with a brief clinical impression. While this standardized format simplifies model training, it significantly hinders the model's flexibility and real-world applicability. As shown in Fig.~\ref{fig:tsne_text}, features derived from free-format text inputs exhibit a much broader distribution in the latent space compared to those from fixed-format inputs. Clinical scenarios are inherently diverse, with varying levels of detail, complexity, and focus depending on the patient and the specific diagnostic needs. This rigidity restricts the model’s ability to generalize across different contexts and user needs. By enabling models to accept a broader range of textual prompts these generative systems could dramatically enhance their adaptability and accessibility. 

In this paper, we introduce \textbf{Text2CT}, a novel framework for high-resolution 3D CT volume generation from free-text descriptions. Text2CT is designed with a modular architecture comprising two specialized 3D networks: a 3D Compression Network and a text-conditional Latent Diffusion Model (LDM)~\cite{Rombach_2022_CVPR}. The proposed method addresses two key limitations in existing approaches, as discussed above: (1) \textbf{high memory consumption} for high-resolution synthesis within a unified 3D network and (2) \textbf{limited generalizability} caused by fixed-format input constraints. The 3D Compression Network plays a crucial role in efficiently handling the large data volumes inherent in high-resolution 3D medical imaging. By compressing 3D CT volumes into a compact latent space, this network significantly reduces memory usage while preserving essential anatomical details. 
We adopt the off-the-shelf 3D Compression Network from MAISI~\cite{guo2024maisi}, which is trained on an extensive dataset of over 55,000 3D medical volumes, ensuring robust performance across diverse anatomical structures and imaging conditions. At the core of Text2CT is the text-conditional LDM, which generates realistic 3D CT volumes based on free-text descriptions. Unlike traditional models that rely on predefined, rigid input formats, we introduce a new prompt formulation that utilizes the capabilities of Large Language Models (LLMs) to transform fixed-format information from radiology reports into diverse and creative textual inputs. To this end, the proposed framework can interpret a wide range of textual prompts, capturing the richness and variability of clinical narratives. Such flexibility would allow healthcare professionals to input more personalized and context-specific information, leading to more accurate and clinically relevant 3D medical image generation. 

To summarize, this paper makes the following contributions:
\begin{enumerate}

  \item We propose the first unified 3D framework effectively addressing the challenges of generating high-resolution 3D CT volumes from free-text descriptions, which integrates a 3D Compression Network with a text-conditional LDM.
  
  \item Text2CT advances beyond fixed-format input by leveraging a new text prompt formulation that utilizes the capabilities of LLM, broadening the model’s applicability in diverse clinical scenarios.

  \item Comprehensive evaluations show Text2CT can significantly outperform baselines in multiple metrics and data augmentation. Human expert assessments demonstrate its superior text alignment and 3D anatomical realism.
\end{enumerate}




%% file: sec/2_related.tex
\section{Related Work}
Recent advancements in generative models, particularly Generative Adversarial Networks (GANs)~\cite{goodfellow2014generative} and Diffusion Models (DMs)~\cite{ho2020denoising}, have revolutionized the ability to generate high-quality, photo-realistic images across various applications in computer vision. These models have been extensively explored for their potential to synthesize realistic visuals in tasks such as image super-resolution~\cite{chen2022real,saharia2022image,lu2022transformer}, style transfer~\cite{jin2022deep,chen2021artistic}, and scene generation~\cite{de2022next,zhang2020deep,gafni2022make}. In the realm of medical imaging, generative models have shown great promise in various applications. For instance, they have been successfully employed for multi-contrast MRI and CT image synthesis~\cite{joyce2017robust,guo2020anatomic,sun2022hierarchical}, enabling the generation of missing imaging modalities or enhancing existing ones. Additionally, cross-modality image translation has allowed for the seamless conversion between different imaging types, such as MRI to CT or vice versa, as demonstrated by several studies~\cite{chartsias2017multimodal,shin2018medical,zhao2017whole,yang2020unsupervised,dewey2019deepharmony}. Beyond synthesis, these models have also contributed to advancements in image reconstruction tasks~\cite{peng2022towards,xie2022measurement,zhao2020smore,darestani2024ir}, where they aid in reconstructing high-quality images from incomplete or noisy data.

\begin{figure*}[ht!] 
\centering
\includegraphics[width=\textwidth]{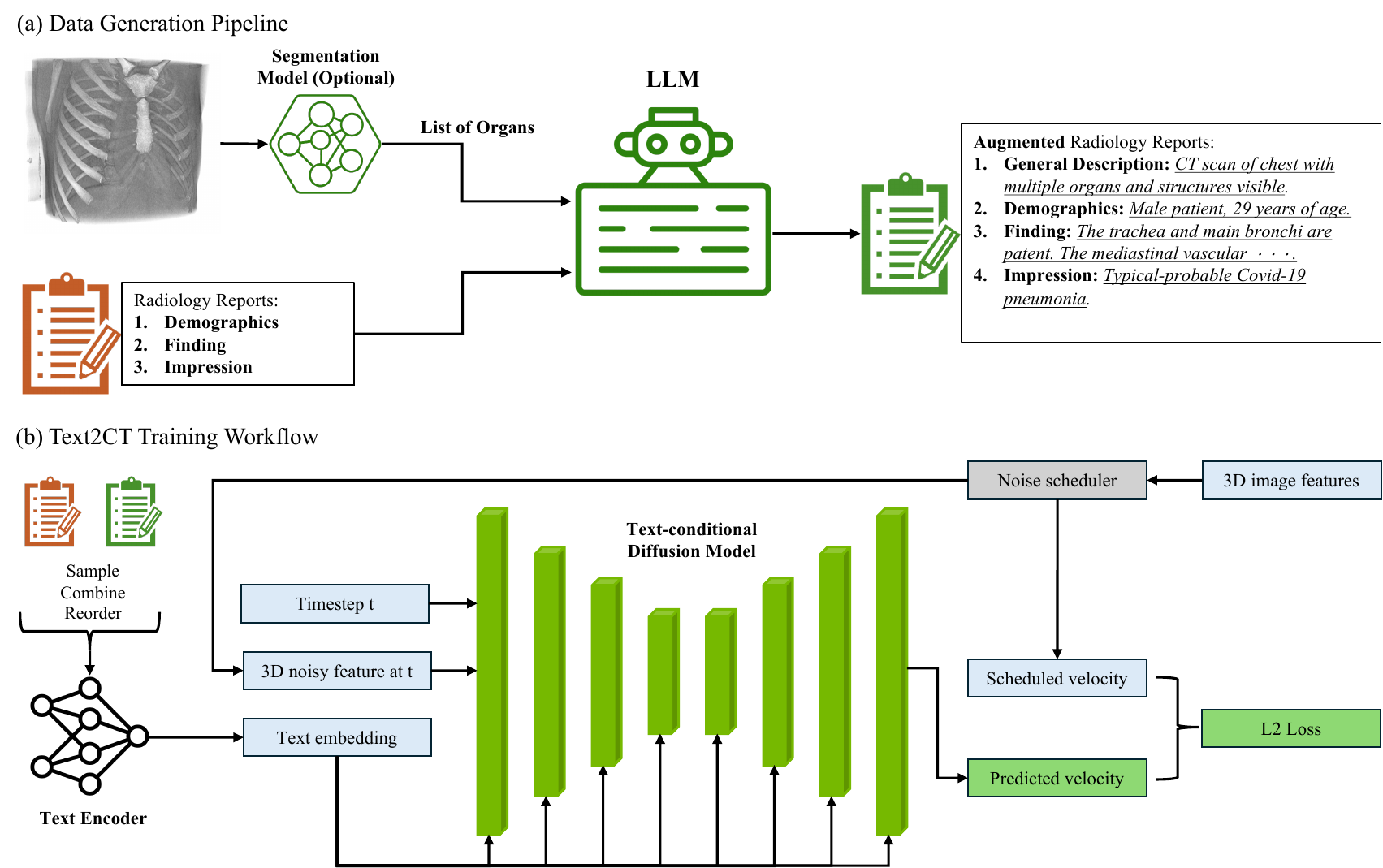}
\caption{(a) The schematics of generation pipelines for Text2CT. We employ LLM to generate the general description and augmented variants of demographics, findings, and impressions based on existing radiology reports and the list of organs derived from segmentation maps. (b) The overview of the training stage of text-conditional LDM in Text2CT.}
\label{fig:workflow}
\end{figure*}

In the field of medical image generation from textual data, studies such as RoentGen~\cite{bluethgen2024vision}, MediSyn~\cite{cho2024medisyn}, MedSyn~\cite{xu2024medsyn}, and GenerateCT~\cite{hamamci2023generatect} stand out as particularly relevant. \cite{bluethgen2024vision,cho2024medisyn} finetunes Stable Diffusion~\cite{Rombach_2022_CVPR} to generate high-quality 2D X-ray images from text prompts with 512$\times$512 pixel resolution. 
MedSyn~\cite{xu2024medsyn} introduces a method for creating 3D lung CT volumes from textual descriptions, employing a hierarchical UNet architecture that initially generates low-resolution images and subsequently enhances them to higher resolutions. However, it is limited to producing volumes with a resolution of 256$^3$ voxels, which falls short of the dimensions typically required for clinical applications. Additionally, its 3D two-stage super-resolution approach~\cite{xu2024medsyn} tends to introduce grid-like artifacts in generated volumes.
GenerateCT~\cite{hamamci2023generatect} integrates a transformer-based generation architecture with a super-resolution diffusion model to produce high-resolution 3D CT scans (e.g., 512$\times$512$\times$201 voxels) from text prompts. Although it achieves promising results in axial views, the use of a 2D super-resolution network introduces issues with 3D continuity, as shown in Fig.~\ref{fig:compare_gct}, which limits its practical application in clinical settings. Moreover, the methods~\cite{xu2024medsyn,hamamci2023generatect} mentioned above restrict input text descriptions to specific formats due to their default prompt formulations. In this work, we aim to enable text-conditioned high-resolution CT volume generation that accommodates free-format text inputs within a unified 3D framework.


%% file: sec/3_method.tex
\section{Methodology}\label{sec:method}

In this section, we first define the problem of text-conditional 3D CT volume generation and then describe our proposed method, Text2CT, in detail. The objective of our approach is to generate a 3D CT volume based on a given text prompt $\mathcal{T}$ by learning a mapping function, $\mathcal{I} = \mathcal{F}(\mathcal{T})$, that translates the text prompt $\mathcal{T}$ into a realistic 3D medical image $\mathcal{I}$. The generated volume should not only be visually realistic but also closely aligned with the descriptive content provided in the text report. As previously discussed, existing methods~\cite{hamamci2023generatect,xu2024medsyn} often rely on fixed-format textual inputs, such as \textit{\{\textbf{age}\} years old \{\textbf{sex}\}:\{\textbf{impression}\}} in GenerateCT~\cite{hamamci2023generatect}, which restricts their accessibility—a critical attribute of any text-conditional generative models. Unlike prior multi-stage generation approaches~\cite{xu2024medsyn,hamamci2023generatect} that apply super-resolution techniques refining output images to desired dimensions, our approach achieves high-resolution text-conditional generation within a unified 3D framework. To the best of our knowledge, this is the first method capable of synthesizing high-resolution 3D CT volumes (512$\times$512$\times$192 voxels) directly from flexible text prompts, significantly enhancing the model's adaptability in diverse clinical applications.

\subsection{Text Prompt Formulation}

In a radiology report, the demographics, findings, and impressions sections each serve specific roles in conveying information about the patient and the results of the imaging study:
\begin{itemize}
  \item \textbf{Demographics} contains patient information such as age, gender, and possibly relevant medical history or symptoms that prompted the imaging study.
  
  \item \textbf{Findings} detail what radiologists observed in the images. This includes descriptions of normal anatomy and any abnormalities or changes from previous scans. The findings are typically descriptive and may include measurements, locations, and characteristics of any lesions or anomalies.

  \item \textbf{Impressions} summarizes the most significant findings and provides an interpretation of what radiologists may indicate in the context of the patient’s health. 
\end{itemize}
The flexibility of text input is crucial for generating clinically relevant 3D CT images. As illustrated in Fig.~\ref{fig:tsne_text}, existing models are typically restricted by rigid text formats, limiting their ability to capture the nuances of diverse clinical scenarios. To address this, Text2CT introduces a new text prompt formulation that trains the proposed text-conditional diffusion model to accommodate free-form clinical descriptions. As illustrated in the data generation pipeline in Fig.~\ref{fig:workflow}(a), each 3D CT volume $\mathcal{I} \in \mathbb{R}^{H\times W \times D}$, represented in grayscale voxel space, where $H$ denotes the height, $W$ the width, and $D$ the
depth, respectively, is paired with a corresponding radiology report $\mathcal{T}$. These reports typically include detailed sections on patient demographics $\mathcal{T}_d$, findings $\mathcal{T}_f$, and impressions $\mathcal{T}_i$, although not all reports include all three sections. A list of organs, denoted as $\mathcal{O}$, is extracted from the CT volume using predicted segmentation masks provided by whole-body CT segmentation models, such as TotalSegmentator~\cite{wasserthal2023totalsegmentator} and VISTA3D~\cite{he2024vista3d}. It is worth noting that Text2CT can use any source providing $\mathcal{O}$ to generate the general description $\mathcal{T}_g$, even eliminating the need for a segmentation model. $\mathcal{O}$ enables accurate identification of specific organs within the volume, which in turn supports the creation of the general description $\mathcal{T}_g$ of the CT scan, succinctly summarizing the body regions and anatomical structures present. This process facilitates more targeted and contextually relevant text-to-image generation by defining the precise areas covered in the scan. Notably, generating general descriptions  $\mathcal{T}_g$ offers the potential to utilize large datasets of CT volumes that lack corresponding radiology reports, but this is orthogonal to this study and will be explored in our future work.

As shown in Fig~\ref{fig:workflow}(a), we leverage the capabilities of the Large Language Model (LLM), such as Llama-3-70B~\cite{dubey2024llama}, to convert structured radiology reports into augmented variants encompassing varied levels of detail and language styles. 
This textual data creation process can be defined as follows:
\begin{equation}
\begin{aligned}
    \mathcal{T}_g, \hat{\mathcal{T}}_d, \hat{\mathcal{T}}_f, \hat{\mathcal{T}}_i = LLM( \mathcal{O},\mathcal{T}_d, \mathcal{T}_f, \mathcal{T}_i),
\end{aligned}
\end{equation}
where $\hat{\mathcal{T}}_d$, $\hat{\mathcal{T}}_f$, and $\hat{\mathcal{T}}_i$ denote the augmented variants of $\mathcal{T}_d$, findings $\mathcal{T}_f$, and impressions $\mathcal{T}_i$, respectively. 
In the training process, we start by randomly selecting $n \in \{1, 2, 3, 4\}$ prompt types from the four aforementioned categories to introduce variability. Then, we shuffle their order to ensure diverse combinations and randomly replace sections of the original text $\mathcal{T}$ with corresponding augmented text $\hat{\mathcal{T}}$. The rationale of the proposed text prompt formulation is to enhance the model’s ability to handle various prompt structures and improve its robustness in generating outputs from diverse textual inputs. Our empirical results suggest that this formulation not only allows Text2CT to interpret rich clinical narratives beyond fixed templates but also substantially enhances both the quality of the generated images and their alignment with the corresponding text descriptions. Finally, the formulated text prompts are processed by a text encoder, which generates the text embedding $\bm{c}_{\text{text}}$ serving as the condition of the diffusion model. We conduct experiments using both biomedical-specific (BiomedCLIP~\cite{zhang2023biomedclip}) and general-purpose (T5~\cite{raffel2020exploring}) text encoders to evaluate their performance in this study.

\subsection{3D Compression Network}

A major challenge in generating high-resolution 3D CT volumes is managing the large memory requirements associated with high-dimensional data. Text2CT addresses this issue with a 3D Compression Network, which encodes high-resolution 3D CT volumes into a compact latent space representation, significantly reducing memory usage while retaining essential anatomical details. Our compression network is adopted from the MAISI~\cite{guo2024maisi} and has been trained on an extensive dataset, ensuring robust generalization across various anatomical regions.

The 3D Compression Network consists of a visual encoder $\mathcal{E}$, which compresses the 3D volume data into a low-dimensional latent representation $z= \mathcal{E} (\mathcal{I})  \in \mathbb{R}^{\frac{H}{4}\times \frac{W}{4} \times \frac{D}{4}}$, and a visual decoder $\mathcal{D}$, which reconstructs the latent representation back to a high-resolution volume $\hat{\mathcal{I}} = \mathcal{D}(z) = \mathcal{D}(\mathcal{E}(\mathcal{I}))$. Generating high-resolution 3D volumes presents significant memory challenges within 3D convolutional networks, often easily exceeding GPU memory limitations. Approaches like super-resolution~\cite{hamamci2023generatect,saharia2022photorealistic} and sliding-window inference~\cite{cardoso2022monai} help but lead to 3D continuity issues or artifacts at boundary transitions, which are especially problematic in image synthesis. To address this, we utilize the tensor splitting parallelism (TSP)~\cite{guo2024maisi} in our 3D Compression Network, which divides large feature maps into smaller segments distributed across multiple GPUs or processed sequentially on a single device, effectively reducing peak memory usage and enhancing inference efficiency. This enables Text2CT to generate and process 3D CT volumes at 512$\times$512$\times$192 voxel resolution, ensuring a seamless flow within a unified 3D framework.

\subsection{Text-conditional Diffusion Model}

The core generation capability of Text2CT lies in the text-conditional LDM~\cite{Rombach_2022_CVPR}, which synthesizes realistic 3D CT volumes conditioned on free-text inputs. This LDM operates within the compact latent space generated by the 3D Compression Network, enabling the efficient generation of anatomically precise 3D images while keeping memory requirements relatively low. Diffusion models are a class of probabilistic models designed to approximate a target data distribution $p(x)$ by progressively refining a noise-distorted variable. These models function by effectively reversing the dynamics of a predefined Markov Chain across a sequence of $T$ steps. A common technique employed in these models is denoising score-matching~\cite{song2020score}, which has gained significant traction in image synthesis applications, as discussed in works~\cite{dhariwal2021diffusion,saharia2022image}.

As shown in Fig~\ref{fig:workflow}(b), we utilize v-prediction~\cite{salimans2022progressive} loss to enhance the performance of the time-conditional U-Net $v_\theta$ as applied in the LDM framework~\cite{Rombach_2022_CVPR}. This U-Net acts as a series of denoising autoencoders, where each step in the sequence is dedicated to predicting a progressively cleaner version of the latent feature $z_t$, a noise-corrupted representation of the original data, over a series of time steps $t$. The architecture of 
$v_\theta$ is designed to adapt this processing based on the specific time step, enabling it to handle the dynamics of noise reduction with high precision throughout the diffusion process. By using v-prediction loss~\cite{salimans2022progressive}, our model predicts the velocity $v$ between noisy and clean states, allowing it to handle the denoising sequence more smoothly. This approach improves stability, sample quality, and training efficiency, making it more effective in reconstructing high-fidelity data through the diffusion process~\cite{salimans2022progressive} and is adopted by many recent leading diffusion models~\cite{podell2023sdxl,videoworldsimulators2024,polyak2025moviegencastmedia}. To process the text embedding $\bm{c}_{\text{text}}$, we leverage the cross-attention conditioning mechanisms~\cite{jaegle2021perceiver} mapping $\bm{c}_{\text{text}}$ to the intermediate representation of time-conditional U-Net $v_\theta$. Based on volume-text pairs, we then learn the
text-conditional LDM via
\begin{equation}
\begin{aligned}
    L_\theta = \mathbb{E}_{\mathcal{E}(\mathcal{I}),t, \bm{c}_{\text{text}}} \Bigl[\lVert v - v_\theta(z_t, t, \bm{c}_{\text{text}}) \rVert_2^2\Bigr],
\end{aligned}
\end{equation}
where the neural backbone $v_\theta$ is configured to condition on time step $t$ and the text conditions as $\bm{c}_{\text{text}}$.

%% file: sec/4_experiments.tex
\section{Experiments}

\subsection{Datasets and Implementation Details}\label{sec: data}
\begin{figure}[t] 
\centering
\includegraphics[width=\columnwidth]{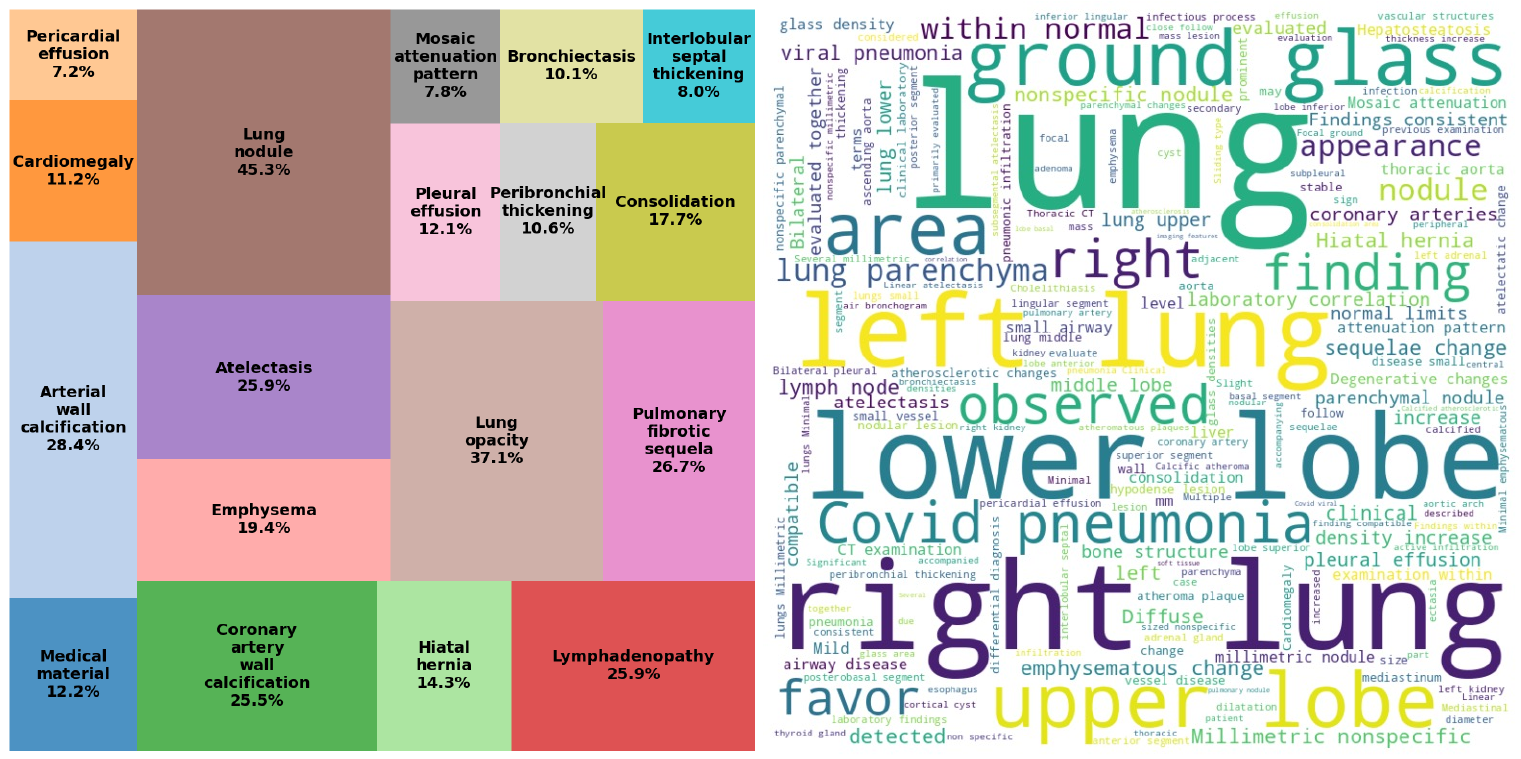}
\caption{The characteristics of the CT-RATE dataset~\cite{hamamci2024foundation} utilized by the proposed Text2CT are detailed through the abnormality distribution (left) and the word cloud of the impression (right).}
\label{fig:dataset}
\end{figure}

We employ two publicly available datasets in our experiments, including CT-RATE~\cite{hamamci2024foundation} and RadChestCT~\cite{draelos2021machine}, that provide high-resolution 3D CT volumes alongside corresponding radiology reports. The primary dataset used for training and validation is derived from CT-RATE~\cite{hamamci2024foundation}, including 50,188 scans for training and 3,039 scans for validation. As shown in Fig.~\ref{fig:dataset}, CT-RATE~\cite{hamamci2024foundation} includes a diverse set of annotated CT volumes covering a wide range of conditions. Additionally, we incorporated RadChestCT~\cite{draelos2021machine} to perform out-of-distribution evaluation and data augmentation experiments using synthetic data. RadChestCT provides 2,286 scans for training and 1,344 scans for validation, offering an effective benchmark for assessing model generalizability beyond the primary dataset. To ensure comprehensive evaluation, we follow~\cite{hamamci2024foundation, hamamci2023generatect} to process all CT volumes to a standardized voxel spacing of 0.75$\times$0.75$\times$1.5 mm and cropped each volume to 512$\times$512$\times$192 voxels, covering a FOV typically encountered in chest CT scans. All networks are implemented using PyTorch\cite{Ansel_PyTorch_2_Faster_2024} and MONAI\cite{cardoso2022monai} frameworks, and model training is conducted on NVIDIA A100 GPUs. 
We evaluate image quality using the Fréchet Inception Distance (FID)~\cite{heusel2017gans} to measure the similarity between the distributions of generated and real images. For assessing text-image alignment, we use the CLIP score, calculated to gauge consistency with text descriptions. As these metrics are 2D-based, we compute them for each volume across three views (axial, sagittal, and coronal). Additional details on data preparation and model training can be found in the supplementary material.

\input{tables/ood_test}

\begin{figure*}[ht!] 
\centering
\includegraphics[width=\textwidth]{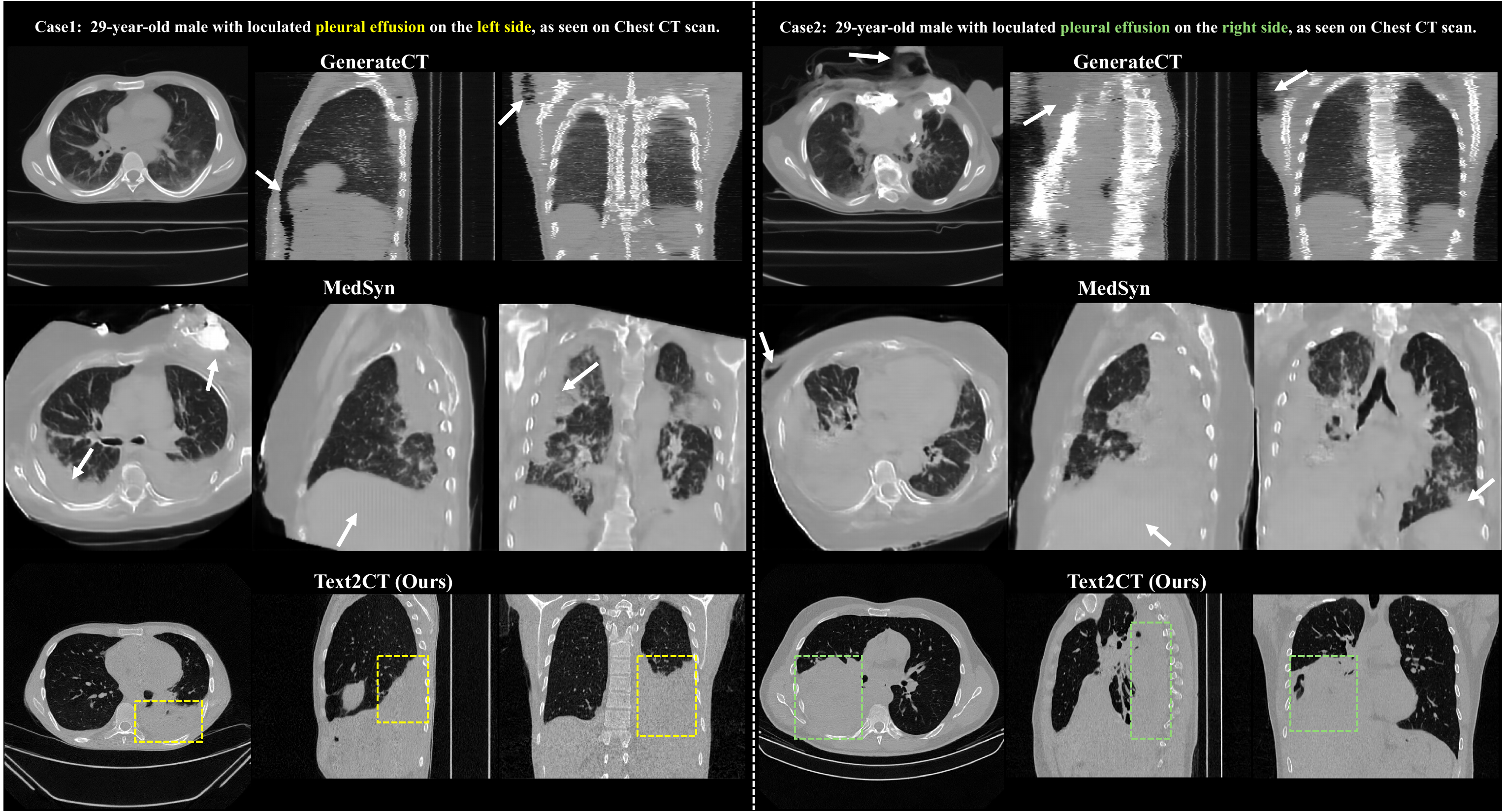}
\caption{\textbf{Qualitative assessment of model generalizability using free-format text prompts.} Abnormalities mentioned in the prompts are highlighted in color and outlined by boxes in generated images. White arrows indicate areas of abnormal anatomy and artifacts from super-resolution in ~\cite{hamamci2023generatect,xu2024medsyn}. Note the \textbf{left-right reversal}: the patient's right side appears on the left side of the image, and vice versa.}
\label{fig:compare_gct_twocases}
\end{figure*}

\subsection{Comparisons with State-of-the-art Methods}
In this section, we benchmark Text2CT against leading medical volume generation models, focusing on synthesis quality, adaptability to free-format text, and clinical value. We explore the potential of using synthetic data for data augmentation. Human expert assessments are also conducted to evaluate the clinical applicability of our approach. 

\noindent\textbf{Out-of-distribution synthesis quality.} To assess the image generation capabilities of Text2CT, we compare its performance with three leading models: MAISI~\cite{guo2024maisi}, GenerateCT~\cite{hamamci2023generatect}, and MedSyn~\cite{xu2024medsyn}. We follow the evaluation setup in~\cite{hamamci2023generatect} that validates models on radiology reports from the RadChestCT dataset~\cite{draelos2021machine}, which has not been used during model training. It is worth noting that while MAISI~\cite{guo2024maisi} does not accept text prompts as inputs, it can be directed to generate chest CT images by controlling the body region prompt. As presented in Table~\ref{tab:ood}, the results show that Text2CT achieves a significantly lower FID and higher CLIP score than three competing methods, indicating superior image quality and more precise text-image alignment. Text2CT’s fully 3D framework maintains a cohesive volumetric structure across slices. This is further supported by the larger metric gap in sagittal and coronal views compared to GenerateCT~\cite{hamamci2023generatect}, emphasizing Text2CT's capability to produce anatomically faithful volumes. Moreover, the superiority of Text2CT's high-resolution synthesis is demonstrated through comparison with MedSyn~\cite{xu2024medsyn} (which generates $256^3$ volumes), as shown by its larger FID gap and smaller CLIP score gap in Table~\ref{tab:ood}.

\input{tables/free_text_gen}
\noindent\textbf{Response to free-format text.} A key requirement for clinical text-to-image models is the ability to accurately interpret diverse, free-format textual descriptions, as radiology reports naturally vary widely in style, detail, and structure. Unlike ~\cite{hamamci2023generatect, xu2024medsyn}, which requires standardized inputs, Text2CT is designed to handle flexible text inputs by leveraging the proposed prompt formation during training. To evaluate the models’ generalizability, we use the LLM~\cite{dubey2024llama} to simulate human-like free-format text prompts and test the model’s generalizability on these varied inputs. Examples of prompts can be found in the supplementary material. As shown in Table~\ref{tab:free_text}, Text2CT consistently achieves better FID and CLIP scores, demonstrating its strong ability to handle variable text prompts. In Fig.~\ref{fig:compare_gct_twocases}, Text2CT successfully captures \textbf{nuanced} variations in the text prompts, such as shifting a lesion location from the left to the right lung. In contrast, GenerateCT~\cite{hamamci2023generatect} and MedSyn~\cite{xu2024medsyn} either generate several anatomically inaccurate outputs or struggle to adhere to textual instructions due to their limited ability to generalize beyond standardized input formats.

\input{tables/data_aug}
\noindent\textbf{Data augmentation by synthetic data.} Given the challenges associated with accessing large quantities of real medical data, we explore the clinical value of using Text2CT-generated synthetic data for data augmentation. The synthetic data is incorporated into a standard training pipeline~\cite{draelos2021machine} for disease classification across 83 types of lung abnormalities. As shown in Table~\ref{tab:data_aug}, adding Text2CT-generated data leads to obvious performance improvements in this task compared to alternatives, with gains of up to 6.2\% in AUROC and 6.9\% in AP. These results highlight the potential of high-quality synthetic data to complement real datasets, providing a scalable solution for expanding training data in medical imaging and enhancing model robustness and generalizability.

\begin{figure}[t] 
\centering
\includegraphics[width=0.70\columnwidth]{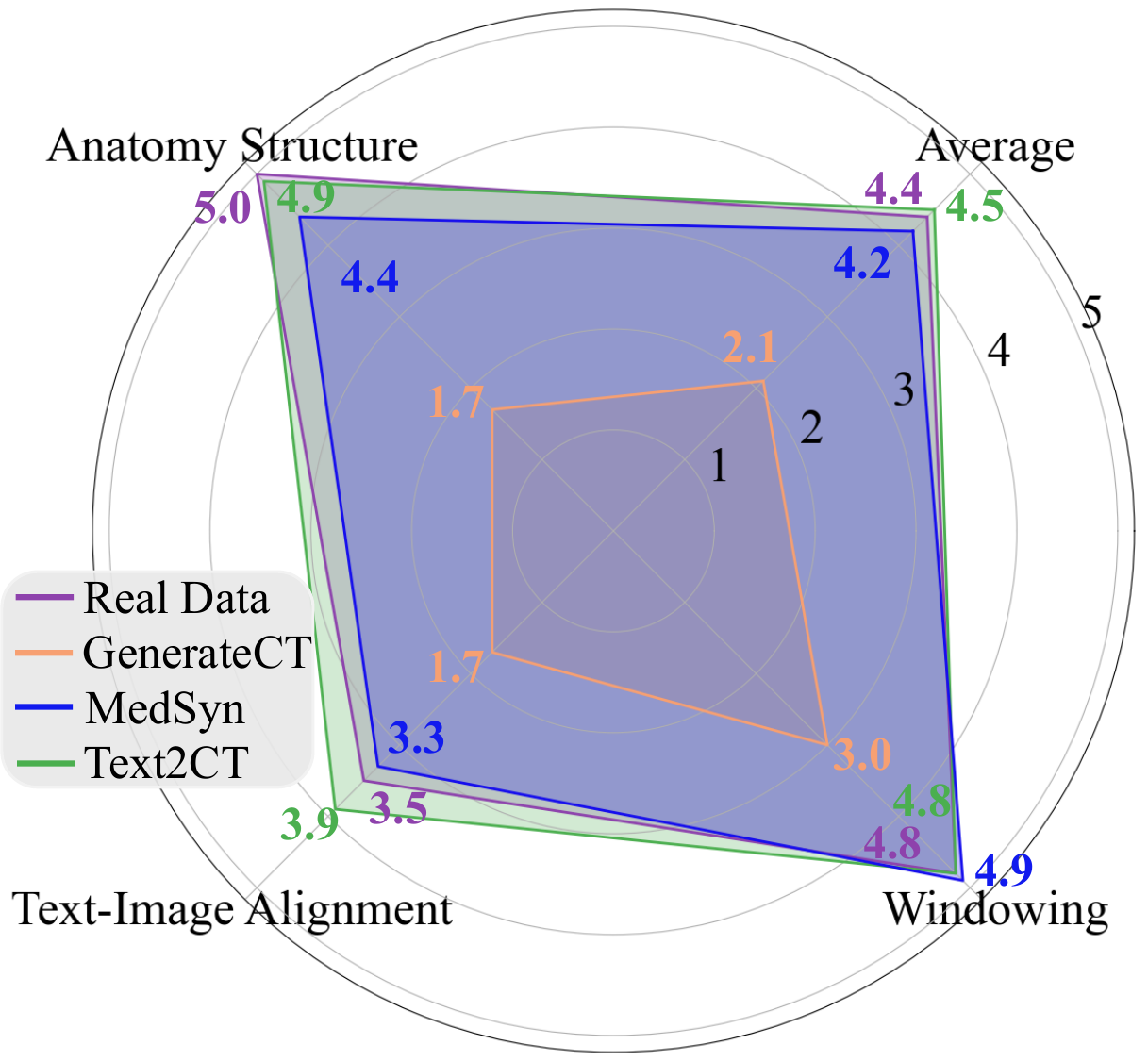}
\caption{\textbf{Human expert assessments of real and synthetic CT volumes.} The radiologist rates each volume on a scale of 1 (poor) to 5 (excellent) in three criteria: Anatomical Structure accuracy, Text-Image Alignment, and Windowing accuracy (\ie, the accuracy of Hounsfield unit across different tissues).}
\label{fig:reader_study}
\end{figure}

\noindent\textbf{Human expert assessments.} To further evaluate the clinical value of generated CT volumes based on free-format text prompts, we conduct a reader study with the radiologist who evaluates both real and synthetic CT volumes. Each volume is rated on a scale from 1 to 5 across three different criteria. As illustrated in Fig.~\ref{fig:reader_study}, Text2CT consistently outperforms GenerateCT~\cite{hamamci2023generatect} and MedSyn~\cite{xu2024medsyn} in evaluation criteria, approaching the ratings of real data. This human expert assessment highlights Text2CT's high anatomical accuracy and strong text-image alignment, underscoring its potential for clinical applications that demand high-quality, anatomically precise images aligned with textual descriptions.

\subsection{Ablation Study}
\input{tables/ab1}
To evaluate the individual contributions of various components within the Text2CT framework, we perform an ablation study that scrutinizes key innovations and variations in input handling. The study particularly focuses on the role of the text encoder and the efficacy of our prompt formulation. As detailed in Table~\ref{tab:ab_1}, we analyze the performance differences between two text encoders, BiomedCLIP~\cite{zhang2023biomedclip} and T5~\cite{raffel2020exploring}, assessing their impact on image quality and text-image alignment on CT-RATE~\cite{hamamci2024foundation} dataset. Each encoder is tested in isolation to measure its effect on the FID and CLIP scores precisely. Moreover, we explore the advantages of our proposed text prompt formulation over the fixed-format prompts. The results clearly indicate that training with our proposed formulation significantly enhances both the visual quality of generated images and their alignment with the text descriptions. Unless otherwise mentioned, the proposed Text2CT referenced is trained using the T5~\cite{raffel2020exploring} text encoder along with our proposed prompt formulation. 
\input{tables/ab2}

In Table~\ref{tab:ab_2}, we assess the quality of images generated by Text2CT based on varying types of input prompt types. Initially, the model processes the general description $\mathcal{T}_g$, and incrementally, more detailed information from all prompt types available in the radiology report is added. We observe a corresponding improvement in image quality with the incremental addition of information to the model. This progression underscores Text2CT's robust response to the enhanced text prompt detail. These results are consistent with the notion from DALL-E 3~\cite{betker2023improving}, which suggests that text-to-image models significantly benefit from highly descriptive texts, such as detailed impressions and findings from radiology reports in this study’s context. Additional experiment details and visualizations are provided in the supplementary material.

%% file: tables/ood_test.tex
\begin{table}[t!]
\caption{\textbf{Out-of-distribution evaluation on RadChestCT~\cite{draelos2021machine}.}}
\small
\resizebox{0.99\columnwidth}{!}{


\begin{tabular}{c|cccc}
\hline\hline
 \multirow{2}{*}{Method}          & \multicolumn{4}{c}{FID $\downarrow$}                                                   \\ \cline{2-5} 
               & Axial          & Sagittal       & \multicolumn{1}{c|}{Coronal}        & Avg.           \\ \hline 
GenerateCT~\cite{hamamci2023generatect}                             & 9.12           & 11.38          & \multicolumn{1}{c|}{17.65}          & 12.71          \\
MedSyn~\cite{xu2024medsyn}                                  & 9.69           & 10.53          & \multicolumn{1}{c|}{11.11}           & 10.44          \\ 
MAISI~\cite{guo2024maisi}                                  & 3.57           & 6.26           & \multicolumn{1}{c|}{9.63}           & 6.49           \\ \hline
\textbf{Text2CT (Ours)}                       & \textbf{2.22}  & \textbf{1.47}  & \multicolumn{1}{c|}{\textbf{1.79}}  & \textbf{1.83}  \\ \hline\hline
\multicolumn{1}{c|}{\multirow{2}{*}{Method}} & \multicolumn{4}{c}{CLIP Score $\uparrow$}                                              \\ \cline{2-5} 
\multicolumn{1}{l|}{}                  & Axial          & Sagittal       & \multicolumn{1}{c|}{Coronal}        & Avg.           \\ \hline
GenerateCT~\cite{hamamci2023generatect}                             & 30.96          & 23.65          & \multicolumn{1}{c|}{24.94}          & 26.52          \\
MedSyn~\cite{xu2024medsyn}                                   & 30.66              & 30.76              & \multicolumn{1}{c|}{27.1}              & 29.52              \\ 
MAISI~\cite{guo2024maisi}                                  & -              & -              & \multicolumn{1}{c|}{-}              & -              \\ \hline
\textbf{Text2CT (Ours)}                       & \textbf{31.16} & \textbf{32.29} & \multicolumn{1}{c|}{\textbf{26.99}} & \textbf{30.15} \\ \hline
\end{tabular}

}
\label{tab:ood}
\end{table}

%% file: tables/free_text_gen.tex
\begin{table}[t!]
\caption{\textbf{Quantitative comparisons of models' generalizability on free-format text from CT-RATE~\cite{hamamci2024foundation}.}}
\small
\resizebox{0.99\columnwidth}{!}{

\begin{tabular}{c|cccc}
\hline\hline
\multirow{2}{*}{Method}                                       & \multicolumn{4}{c}{FID $\downarrow$}                                                   \\ \cline{2-5} 
                                                              & Axial          & Sagittal       & \multicolumn{1}{c|}{Coronal}        & Avg.           \\ \hline
GenerateCT~\cite{hamamci2023generatect} & 10.50          & 13.25          & \multicolumn{1}{c|}{19.14}          & 14.30          \\
MedSyn~\cite{xu2024medsyn} & 10.03          & 10.06          & \multicolumn{1}{c|}{10.54}          & 10.21          \\ \hline
\textbf{Text2CT (Ours)}                                       & \textbf{0.85}  & \textbf{0.67}  & \multicolumn{1}{c|}{\textbf{0.82}}  & \textbf{0.78}  \\ \hline\hline
\multicolumn{1}{c|}{\multirow{2}{*}{Method}}                        & \multicolumn{4}{c}{CLIP Score $\uparrow$}                                              \\ \cline{2-5} 
\multicolumn{1}{l|}{}                                         & Axial          & Sagittal       & \multicolumn{1}{c|}{Coronal}        & Avg.           \\ \hline
GenerateCT~\cite{hamamci2023generatect} & 26.79          & 21.68          & \multicolumn{1}{c|}{21.50}          & 23.32        \\
MedSyn~\cite{xu2024medsyn} & 28.43          & 24.53         & \multicolumn{1}{c|}{21.45}          & 24.81          \\ \hline
\textbf{Text2CT (Ours)}                                       & \textbf{30.58} & \textbf{29.84} & \multicolumn{1}{c|}{\textbf{24.34}} & \textbf{28.25} \\ \hline
\end{tabular}


}
\label{tab:free_text}
\end{table}

%% file: tables/data_aug.tex
\begin{table}[t!]
\caption{\textbf{Data augmentation experiments on RadChestCT~\cite{draelos2021machine}.}}
\small
\resizebox{0.99\columnwidth}{!}{
\begin{tabular}{cccc}
\hline\hline
Training Data                    & Method                  & AUROC $\uparrow$          & AP $\uparrow$             \\ \hline
Real Only                   & -                       & 0.613          & 0.177          \\ \hline
\multirow{3}{*}{Syn Only}    & GenerateCT~\cite{hamamci2023generatect}              & 0.536 (\textcolor{Red}{-7.7\%})         & 0.146 (\textcolor{Red}{-3.1\%})         \\ 
& MedSyn~\cite{xu2024medsyn}             & 0.559 (\textcolor{Red}{-5.4\%})         & 0.158 (\textcolor{Red}{-1.9\%})         \\ \cline{2-4} 
                                 & \textbf{Text2CT (Ours)} & \textbf{0.586} (\textcolor{Red}{-2.7\%}) & \textbf{0.168} (\textcolor{Red}{-0.9\%}) \\ \hline
\multirow{3}{*}{Syn + Real} 
& GenerateCT~\cite{hamamci2023generatect}              & 0.623 (\textcolor{Green}{+1.0\%})         & 0.190 (\textcolor{Green}{+1.3\%})          \\
& MedSyn~\cite{xu2024medsyn}              & 0.647 (\textcolor{Green}{+3.4\%})         & 0.202 (\textcolor{Green}{+2.5\%})          \\ \cline{2-4} 
                                 & \textbf{Text2CT (Ours)} & \textbf{0.675} (\textcolor{Green}{+6.2\%})  & \textbf{0.246} (\textcolor{Green}{+6.9\%})  \\ \hline
\end{tabular}

}
\label{tab:data_aug}
\end{table}

%% file: tables/ab1.tex
\begin{table}[t!]
\caption{\textbf{Ablation study of proposed text prompt formulation and text encoders on CT-RATE~\cite{hamamci2024foundation}.}}
\resizebox{0.99\columnwidth}{!}{
\begin{tabular}{c|c|c|c}
\hline\hline
Text Encoder & Prompt Formulation & FID (Avg.) $\downarrow$ & CLIP Score (Avg.) $\uparrow$ \\ \hline
BiomedCLIP   & Fix-format         & 1.98                    & 26.59                        \\
T5-Base    & Fix-format         & 2.10                    & 26.52                        \\ \hline
BiomedCLIP   & \textbf{Proposed}           & 0.83                    & \textbf{27.93}               \\
T5-Base    & \textbf{Proposed}             & \textbf{0.65}           & 27.82                        \\ \hline
\end{tabular}
}
\label{tab:ab_1}
\end{table}

%% file: tables/ab2.tex
\begin{table}[t!]
\caption{\textbf{Ablation study on variants of input prompt types on CT-RATE~\cite{hamamci2024foundation}.}}
\small
\resizebox{0.99\columnwidth}{!}{

\begin{tabular}{c|cccc}
\hline\hline
\multirow{2}{*}{Input prompt types}      & \multicolumn{4}{c}{FID $\downarrow$}                            \\ \cline{2-5} 
                                   & Axial & Sagittal & \multicolumn{1}{c|}{Coronal} & Avg.          \\ \hline
$\mathcal{T}_g$                              & 2.59  & 1.84     & \multicolumn{1}{c|}{2.20}    & 2.21          \\
$\mathcal{T}_g$  , $\mathcal{T}_i$                    & 0.99  & 0.62     & \multicolumn{1}{c|}{0.76}    & 0.79          \\
$\mathcal{T}_d$ , $\mathcal{T}_i$                   & 0.67  & 0.54     & \multicolumn{1}{c|}{0.74}    & 0.65          \\
$\mathcal{T}_d$, $\mathcal{T}_i$, $\mathcal{T}_f$         & 0.64  & 0.49     & \multicolumn{1}{c|}{0.71}    & 0.61          \\
$\mathcal{T}_g$ , $\mathcal{T}_f$, $\mathcal{T}_i$, $\mathcal{T}_f$ & \textbf{0.55}  & \textbf{0.48}     & \multicolumn{1}{c|}{\textbf{0.70}}    & \textbf{0.58} \\ \hline
\end{tabular}

}
\label{tab:ab_2}
\end{table}

%% file: sec/5_discussion.tex
\section{Discussion and Limitations}
The proposed Text2CT model demonstrates notable advancements over prior methods in synthesis quality, text-image alignment, and adaptability to free-format text descriptions. However, some limitations require further exploration. The model performance depends on the detail and quality of the input text. As shown in Table~\ref{tab:ab_2}, less descriptive texts could lead to less accurate image generation. Effectively leveraging generated general descriptions, $\mathcal{T}_g$, derived from CT volumes without corresponding radiology reports is a promising direction for future work. Moreover, despite optimization efforts, the high computational demands associated with generating 3D volumes may restrict its use in resource-constrained settings. Future work could focus on memory-efficient methods to address these requirements, ultimately broadening Text2CT’s accessibility for diverse clinical applications.

%% file: sec/6_conclusion.tex
\section{Conclusion}

This study presents Text2CT, a 3D text-conditional generative model that synthesizes high-resolution, anatomically accurate CT volumes conditioned on flexible clinical text prompts. Text2CT marks a step forward in generating clinically applicable CT volumes from unstructured text descriptions. With robust text interpretation, and effective data augmentation capabilities, Text2CT offers a scalable solution for overcoming data limitations and enhancing model performance across medical imaging applications.

%% file: sec/7_suppl.tex
\clearpage
\setcounter{page}{1}
\maketitlesupplementary
\appendix
\beginsupplement

This supplementary material is organized as follows: More implantation details about two networks, computational details, and evaluation metrics are provided in Sec.~\ref{supp:imp}. Sec.~\ref{supp:exp} contains additional ablation studies, example prompts for LLM, and visualizations of synthetic data.

\section{Additional Implementation Details}\label{supp:imp}

\subsection{3D Compression Network}

Extensive data augmentations are applied to CT and MR images to train the 3D Compression Network as a foundational model. CT intensities are clipped to a Hounsfield Unit range of -1000 to 1000 and normalized to [0,1], while MR intensities are scaled to [0,1] using the 0th to 99.5th percentile range and enhanced with augmentations like random bias fields, Gibbs noise, contrast adjustments, and histogram shifts. Both modalities underwent spatial augmentations, including random flipping, rotation, intensity scaling, shifting, and resolution changes. We adopt the off-the-shelf 3D Compression Network from MAISI~\cite{guo2024maisi} VAE, which is trained in two phases: first, for 100 epochs on [64,64,64] patches to enhance generalization for partial volume effects, and then for 200 epochs on [128,128,128] patches to capture richer contextual information, improving overall performance. More details about the training procedure and datasets can be found in the MAISI official GitHub Repo\footnote{https://monai.io/research/maisi}.

\subsection{Text-conditional Diffusion Model}~\label{supp:dm}
The Text2CT diffusion model training involves a series of data preprocessing for optimal learning. The process begins with loading the images, ensuring the correct channel structure, orienting them according to the ``RAS" axcode, and normalizing intensity values by scaling them from -1000 to 1000 into the range [0,1]. Images are then resampled to a standardized voxel spacing of 0.75$\times$0.75$\times$1.5 mm and cropped each volume to 512$\times$512$\times$192 using trilinear interpolation, with spatial details recorded. Each preprocessed image is passed through a pre-trained 3D Compression Network to generate a compressed latent representation, which is stored for subsequent training. Additional input attributes required by the diffusion model, such as text prompts, are extracted from the corresponding radiology report. The list of organs in each volume is extracted from segmentation maps using segmentation tools like TotalSegmentator~\cite{wasserthal2023totalsegmentator}. 

The Text2CT diffusion model training begins with an initial learning rate of $1\times 10^{-4}$, a batch size of 4, and spans 500 epochs. Training employs a U-Net architecture for velocity prediction~\cite{salimans2022progressive}, leveraging distributed computing for efficiency on multiple GPUs (32 A100 GPUs). An Adam optimizer updates model parameters, while a polynomial learning rate scheduler adjusts the learning rate across training steps. Noise is systematically added to the data via a noise scheduler, and the model iteratively reduces this noise using an L2 loss function. Mixed precision training and gradient scaling are used to optimize memory and computational efficiency, ensuring the model can handle the complexity of the data and training process effectively.

\input{tables/supp_compute}
\subsection{Computational Details}
The training and evaluation of the Text2CT diffusion model were performed on a high-performance computing cluster utilizing 32 NVIDIA A100 GPUs, each with 80GB of memory, for distributed processing. During training, processing a single batch of data (batch size of 1) required approximately 0.3 seconds, with a peak GPU memory usage of 19.4 GB, as shown in Table~\ref{tab:compute}. As discussed in Sec.~\ref{supp:dm}, we precompute the compressed latent representation using the 3D Compression Network during data preprocessing, so we can directly use stored latent representation without the overhead of encoding in diffusion model training. For inference, generating a 512$\times$512$\times$192 volume without optimization techniques takes approximately 145.7 seconds with classifier-free guidance enabled and 218.7 seconds without it. The peak GPU memory usage is dominated by the decoding process of the 3D compression network during inference. We adopt the default setting of MAISI~\cite{guo2024maisi} VAE that leverages tensor splitting parallelism dividing large feature maps into smaller segments (\eg, 8 segments) distributed across multiple GPUs or processed sequentially on a single device, effectively reducing peak memory usage. The increased inference time when using classifier-free guidance is primarily due to the need to perform inference for both conditional and unconditional generation simultaneously in the text-conditional diffusion model, which is equivalent to doubling the inference batch size. 

\subsection{Evaluation Metrics}

The models' performance is evaluated using various quantitative metrics to ensure thorough validation. For image quality in generation tasks, the Fréchet Inception Distance (FID) is employed to measure fidelity and perceptual quality by comparing generated images to real ones. Our FID implementation leverages the \textit{FIDMetric} class from MONAI\footnote{https://github.com/Project-MONAI/MONAI} and uses RadImageNet~\cite{ryai.210315} as the feature extractor. For text-conditional tasks, the alignment between generated images and corresponding text prompts is assessed using embedding-based similarity metrics like the CLIP score, which evaluates the semantic consistency of text-image pairs. Our implementation is based on the \textit{CLIPScore} class from TorchMetrics\footnote{https://github.com/Lightning-AI/torchmetrics}, with BiomedCLIP~\cite{zhang2023biomedclip} serving as the text and image encoder to extract relevant features. Together, these metrics provide a comprehensive evaluation of the model's ability to generate high-quality, semantically aligned, and anatomically accurate outputs.

\section{Supplementary Experiment Results}\label{supp:exp}
\input{tables/supp_cfg}
\subsection{Classifier-free Guidance Scales}\label{supp:cfg}
An additional ablation study, summarized in Table~\ref{tab:cfg}, explores the effect of different classifier-free guidance (CFG) scales on FID and CLIP scores across axial, sagittal, and coronal orientations. The results indicate that increasing the CFG scale improves FID, with the best average score of 0.62 observed at a CFG scale of 7, reflecting superior image quality. Conversely, the CLIP score, which evaluates semantic alignment between text and images, peaks at a CFG scale of 5 with an average score of 27.82, indicating optimal text-image consistency. These findings highlight a trade-off between image fidelity and semantic alignment: a CFG scale of 7 achieves the highest image quality, while a scale of 5 provides the best text-conditional performance. Balancing these factors, the proposed Text2CT framework uses a CFG scale of 5 for inference unless otherwise specified.

\subsection{Example Prompts for LLM}\label{supp:prompt}
\begin{figure}[ht!] 
\centering
\includegraphics[width=\columnwidth]{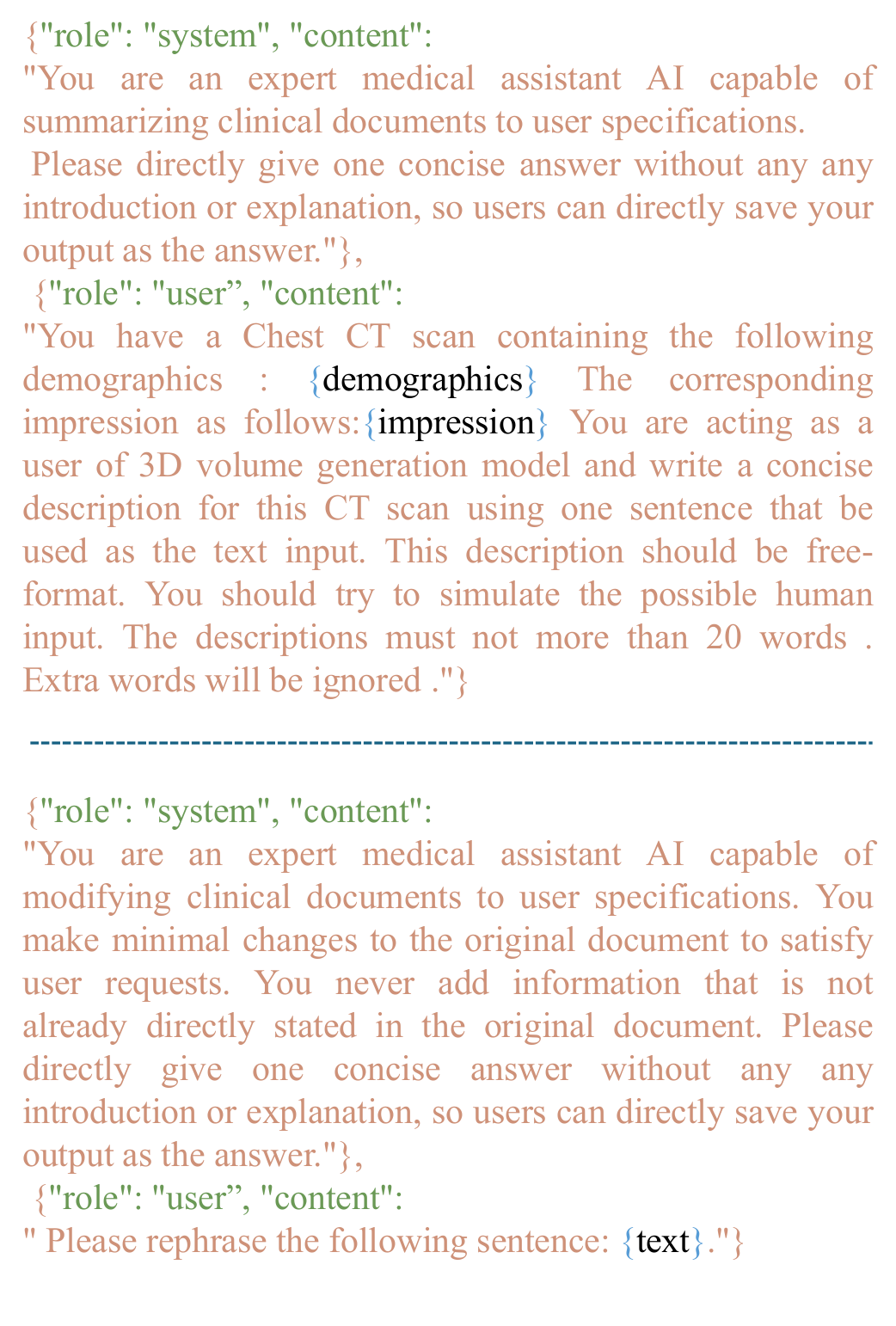}
\caption{\textbf{Example prompts used for text prompt data generation.}. \textit{Top row}: A sample prompt provided to LLM for generating concise free-form text based on demographics and impressions.
\textit{Bottom row}: A sample prompt used to guide the LLM in creating an augmented variation of an existing text.
}
\label{fig:llm_prompt}
\end{figure}

\begin{figure*}[ht!] 
\centering
\includegraphics[width=\textwidth]{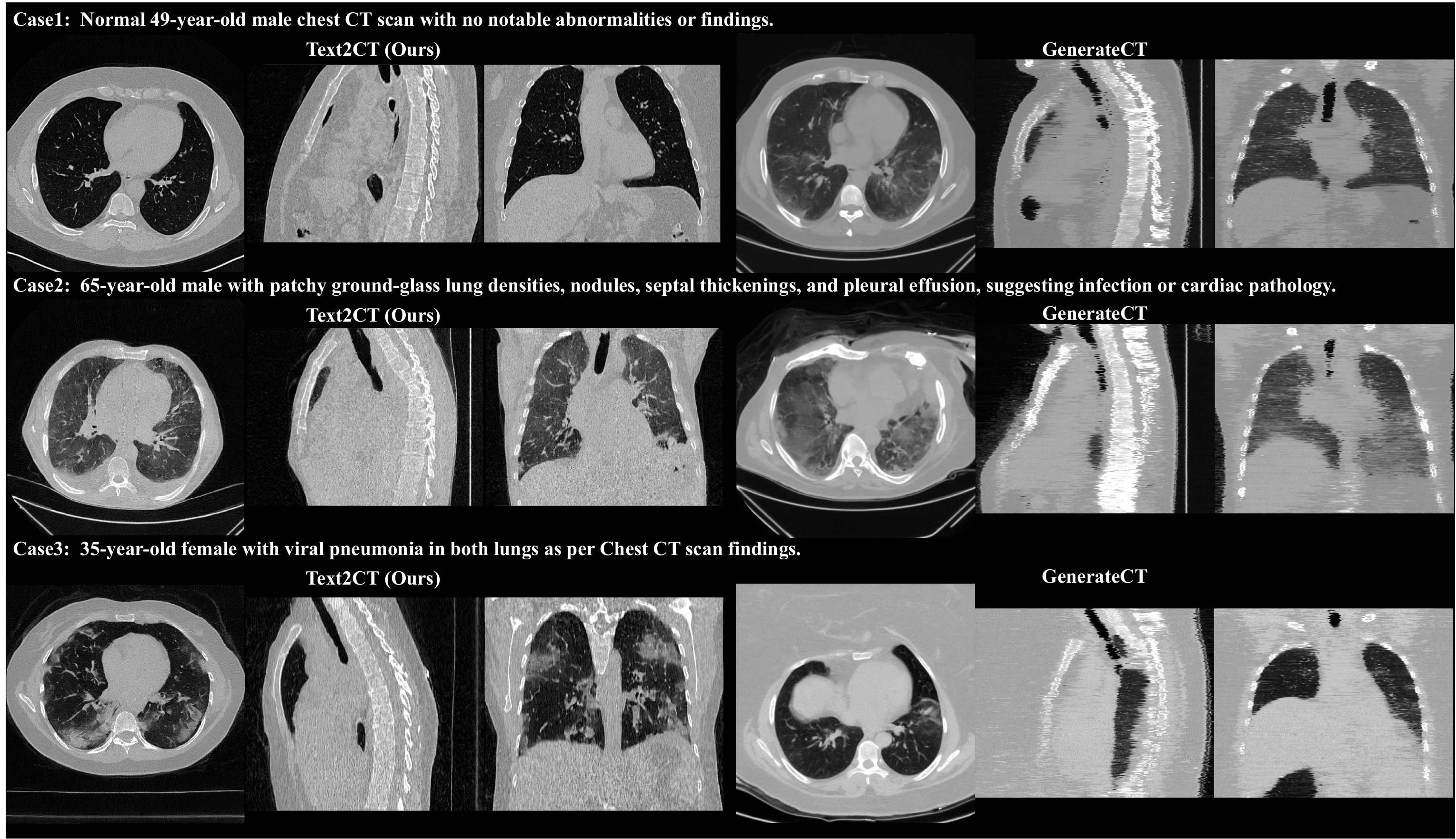}
\caption{\textbf{Additional qualitative assessment of model generalizability using free-format text prompts.}}
\label{fig:supp_compare_gct_twocases}
\end{figure*}

To generate text prompts for training and evaluation, we employ specific examples designed to guide the language model (LLM) in producing the desired outputs. The prompts mainly serve two purposes. The first function involves free-form Text Generation as shown in the top row of Figure~\ref{fig:llm_prompt}, these prompts are used primarily for evaluating the model's generalizability. The LLM is instructed to generate concise free-form text based on patient demographics and corresponding impressions. This ensures the generated text is coherent, relevant, and diverse, enabling a robust assessment of the model's ability to generalize across different textual inputs. The second function pertains to text augmentation, illustrated in the bottom row of Figure~\ref{fig:llm_prompt}, these prompts are designed for augmenting existing text data and are primarily used for model training. The LLM modifies or rephrases the provided input text while maintaining its original semantic meaning. This approach expands the training dataset and improves the model's ability to handle variations in text while preserving the alignment between textual and visual data.

\subsection{Additional Visualizations of Synthetic Data}\label{supp:vis}

Figure~\ref{fig:supp_compare_gct_twocases} illustrates additional qualitative results generated by the Text2CT model, showcasing its capability to produce synthetic CT volumes based on diverse text prompts. These prompts describe various clinical conditions, including healthy lungs, ground-glass opacities, pleural effusion, and pneumonia. The visualizations highlight the model's ability to generate anatomically precise and contextually accurate CT volumes, effectively aligning the visual data with the specific clinical descriptions provided in the text inputs.

%% file: tables/supp_compute.tex
\begin{table}[t!]
\caption{\textbf{The computation details of Text2CT on A100 GPU.} Time (AE) and Time (DM) denote the processing time of one batch of data for the 3D compression network and text-conditional diffusion model, respectively. CFG denotes the classifier-free guidance during inference.}
\small
\resizebox{0.99\columnwidth}{!}{
\begin{tabular}{cccc}
\hline
                  & Time (AE) & Time (DM) & Peak GPU memory \\ \hline
Training          & -                   & 0.3 s               & 19.4 GB          \\ \hline
Inference w/o CFG & 33.7 s              & 112.0 s             & 29.8 GB          \\ \hline
Inference w/ CFG  & 33.7 s              & 185.0 s             & 29.8 GB          \\ \hline
\end{tabular}

}
\label{tab:compute}
\end{table}

%% file: tables/supp_cfg.tex
\begin{table}[t!]
\caption{\textbf{Ablation study on different classifier-free guidance scales.} }
\small
\resizebox{0.99\columnwidth}{!}{
\begin{tabular}{c|cccc}
\hline
\multirow{2}{*}{CFG Scale} & \multicolumn{4}{c}{FID $\downarrow$}                                                   \\ \cline{2-5} 
                           & Axial          & Sagittal       & \multicolumn{1}{c|}{Coronal}        & Avg.           \\ \hline
1                          & 0.89           & 0.64           & \multicolumn{1}{c|}{0.82}           & 0.78           \\
3                          & 0.76           & 0.64           & \multicolumn{1}{c|}{0.83}           & 0.74           \\
5                          & 0.67           & 0.54           & \multicolumn{1}{c|}{0.74}           & 0.65           \\
\textbf{7}                 & \textbf{0.65}  & \textbf{0.52}  & \multicolumn{1}{c|}{\textbf{0.70}}  & \textbf{0.62}  \\ \hline
\multirow{2}{*}{CFG Scale} & \multicolumn{4}{c}{CLIP Score $\uparrow$}                                              \\ \cline{2-5} 
                           & Axial          & Sagittal       & \multicolumn{1}{c|}{Coronal}        & Avg.           \\ \hline
1                          & 29.60          & 29.30          & \multicolumn{1}{c|}{23.91}          & 27.60          \\
3                          & 30.07          & 29.35          & \multicolumn{1}{c|}{23.87}          & 27.76          \\
\textbf{5}                 & \textbf{30.16} & \textbf{29.37} & \multicolumn{1}{c|}{\textbf{23.91}} & \textbf{27.82} \\
7                          & 30.11          & 29.36          & \multicolumn{1}{c|}{23.74}          & 27.74          \\ \hline
\end{tabular}
}
\label{tab:cfg}
\end{table}